\documentclass[fleqn,referee]{envauth}

\received{00 Month 2012}
\revised{00 Month 2012}
\accepted{00 Month 2012}

\usepackage{natbib,upgreek,enumerate}
\usepackage{algorithm2e}
\usepackage{rotating}
\allowdisplaybreaks

\runninghead{L. Sun, C. Lee, and J. A. Hoeting}{Parameter inference and model selection via ABC}

\begin{document}

\title{Parameter inference and model selection in deterministic and stochastic dynamical models via approximate Bayesian computation: modeling a wildlife epidemic}

\author{Libo Sun\corrauth, Chihoon Lee, and Jennifer A. Hoeting}

\corraddr{Libo Sun, E-mail: sun@stat.colostate.edu}

\address{Department of Statistics, Colorado State University, Fort Collins, Colorado, USA}

\begin{abstract}
We consider the problem of selecting deterministic or stochastic models for a biological, ecological, or environmental dynamical process. In most cases, one prefers either deterministic or stochastic models as candidate models based on experience or subjective judgment. Due to the complex or intractable likelihood in most dynamical models, likelihood-based approaches for model selection are not suitable. We use approximate Bayesian computation for parameter estimation and model selection to gain further understanding of the dynamics of two epidemics of chronic wasting disease in mule deer. The main novel contribution of this work is that under a hierarchical model framework we compare three types of dynamical models: ordinary differential equation, continuous time Markov chain, and stochastic differential equation models. To our knowledge model selection between these types of models has not appeared previously. Since the practice of incorporating dynamical models into data models is becoming more common, the proposed approach may be very useful in a variety of applications.
\end{abstract}

\maketitle

\section{Introduction}
In the study of a biological, ecological, or environmental dynamical process, the choice of underlying dynamical model (also known as the \emph{process} model) is usually based upon expert knowledge or non-generalizable, ad hoc preference. Moreover, it is often the case that parameters of the model are not estimated using statistical functions of observed data. The objectives of this paper are (a) \emph{to investigate a systematic statistical approach to select a process model that is consistent with the observed data} and (b) \emph{to produce parameter estimates and quantify associated uncertainties based on the observed data}. We undertake these goals under a hierarchical model framework and demonstrate our approach using ecological models for the transmission of chronic wasting disease (CWD) in mule deer.

In general, the hierarchical model \citep{berliner1996hierarchical,wikle2003hierarchical} consists of three levels: a data model, a process model, and a parameter model. The data model represents measurement error in the observed data, which is very common in epidemiology, ecology and environmental science. For example, the number of deaths due to CWD in a wild population is subject to CWD test accuracy and the expense of data collection. The process model is the scientific model based on theories and simplifications of reality. Deterministic or stochastic models may be adopted as the process model. The parameter model acknowledges parameter uncertainty.

With regards to the \emph{process} model, there could be several candidate models. For instance, in understanding the dynamics of infectious diseases in biology, ecology and environmental science, scientists can adopt a set of ordinary differential equations (ODEs) or a set of stochastic differential equations (SDEs), or a continuous time Markov chain (CTMC). A notable example is the Susceptible-Infected-Removed (SIR) model, which is a commonly used dynamical model \citep{anderson1992infectious,hethcote2000mathematics} in the study of disease transmission (see also \cite{allen2003introduction}). \citet{miller2006dynamics} proposed several ODE models to describe the transmission mechanism of CWD, a fatal contagious disease in cervid populations. Subsequently, an SDE model was proposed by \citet{sun2015penalized} to provide more realistic description of the transmission process of CWD. There are pros and cons of those models; for example, stochastic process models allow process error but deterministic models do not. Due to their simplicity, deterministic dynamical models are typically preferred when studying a large community. Stochastic models define the probability of disease transmission between two individuals, while deterministic models describe the spread under the assumption of mass action. However for a specific dataset, the choice between deterministic or stochastic dynamical models is often subjective. Therefore, a \emph{data-driven} approach to select between the deterministic and stochastic models based on the observed data is needed.

In many contexts model selection is typically performed via a likelihood ratio test, the Akaike information criterion or the Bayesian information criterion. However, such approaches are not suitable for the dynamical models  that are often used in biology and ecology because the likelihood is intractable. Approximate Bayesian computation (ABC) is a methodology to estimate the model parameters when the likelihood is difficult to compute. A simulation-based procedure and a distance function between simulated data and the observed data are used instead of the likelihood in ABC. Various ABC algorithms have been proposed, such as rejection based ABC \citep{pritchard1999population}, regression based ABC \citep{beaumont2002approximate}, and ABC Markov chain Monte Carlo (MCMC) \citep{marjoram2003markov}. \citet{toni2009approximate} developed an ABC method based on sequential Monte Carlo (SMC) \citep{del2006sequential} for parameter estimation and model selection for dynamical models. This ABC SMC algorithm addresses a potential drawback of previous ABC algorithms, such as slow convergence rate, by sampling from a sequence of intermediate distributions. 
\citet{beaumont2010approximate} provides a detailed review of ABC methods.

In this work, we incorporate the ABC SMC algorithm into a hierarchical model framework, and perform parameter estimation (with credible intervals) and dynamical model selection among a set of ODEs, SDEs, and CTMC that arise as models for the transmission of CWD.  To our knowledge model selection between these types of models has not appeared previously. Since the practice of incorporating dynamical models into data models (i.e., a hierarchical framework) is becoming more common, the proposed approach may be useful in a variety of applications.

The remainder of the paper is organized as follows. We provide a brief introduction to CWD in Section \ref{CWD} and present the related hierarchical model framework used to investigate the transmission of CWD in Section \ref{models}. Section \ref{ABC} briefly describes the ABC SMC algorithm in \citet{toni2009approximate}. Section \ref{simulation} presents the performance of the ABC SMC algorithm on simulated datasets. Section \ref{app results} shows the results based on data from two CWD epidemics.  Section \ref{conclusion} concludes with a discussion.

\section{Chronic wasting disease (CWD)}\label{CWD}
Deer populations and ecosystems can be severely disrupted by the contagious prion disease, known as CWD \citep{miller2006dynamics}. Deer populations in many U.S. states are intensely monitored due to hunting. Because of the impact of CWD on the number of deer, it is important to understand the transmission mechanisms of CWD. Several deterministic epidemic models were proposed by \citet{miller2006dynamics} in order to portray the transmission of CWD. Here, based on those deterministic models, we derive CTMC and SDE models for CWD using the techniques described in \citet[Chapter~8]{allen2003introduction}. Then, we implement the ABC SMC approach to the dataset studied in \citet{miller2006dynamics}. Their dataset consists of annual observations of cumulative mortality from two distinct CWD epidemics (Figure \ref{CWD_indirect_fig} upper display) in captive mule deer held at the Colorado Division of Wildlife Foothills Wildlife Research Facility in Fort Collins, Colorado. The first epidemic occurred from 1974 to 1985 and the second epidemic occurred in a new deer herd from 1992 to 2001. The dataset also includes the annual number of new deer added to the herd and the per capita losses due to natural deaths and removals. We assume key model parameters, such as the direct transmission coefficient $\beta$, the per capita CWD mortality rate $\mu$, the indirect transmission coefficient $\gamma$, the per capita rate of excretion of infectious material by infected animals $\epsilon$, and the mass-specific rate of loss of infectious material from the environment $\tau$, are innate characteristics of the population and the associated disease and do not change between these two epidemics. Biologists with considerable expertise in CWD have previously made the same assumption \citep{miller2006dynamics}. Moreover, it is not possible to get accurate parameter  estimates if you consider the two epidemics separately for such a small sample size. 

\section{Hierarchical model framework}\label{models}
A hierarchical model is a natural choice for many problems in ecology because there are typically multiple sources of uncertainty  \citep{berliner1996hierarchical,wikle2003hierarchical}. There are three stages in the hierarchical model framework:
\begin{description}
\item[Data Model:] Specify the distribution of the data given the process model.
\item[Process Model:] Describe the process conditional on process parameters.
\item[Parameter Model:] Account for uncertainty in the process parameters.
\end{description}
Below we develop several hierarchical models for the CWD data.

\subsection{Data model}\label{data_model}
To allow for measurement and observation error in the observed counts, we consider two possible data models for the transmission of CWD.  At time $t$ let  $S(t)$ denote the number of susceptible animals, $I(t)$ denote the number of infected animals, $C(t)$ denote the true unobserved number of accumulated deaths from CWD, and $\tilde{C}(t)$ denote the \emph{observed} accumulated CWD deaths. We assume that only $\tilde{C}(t)$ is observed at discrete time $t=t_0, t_1,\dots,t_n$, and is modeled by
\begin{equation}\label{data_binom}
\tilde{C}(t)\sim\text{Binomial}\left(N(t);\frac{C(t)}{N(t)}\right),
\end{equation}
where $N(t)=S(t)+I(t)+C(t)$ is the total number of animals (including deaths) at time $t$.  As an alternative data model we also consider
\begin{equation}\label{data_pois}
\tilde{C}(t)\sim\text{Poisson}\left(C(t)\right).
\end{equation}
Note that this model allows for the case where the observed number of animals at time $t$ $\tilde{C}(t)$ to be larger than the total number of animals $N(t)$. When such an assumption is not reasonable, it is necessary to constrain $\tilde{C}(t)\leq N(t)$. Without loss of generality, we assume $C(t_0)=\tilde{C}(t_0)=0$.

\subsection{Process model}\label{Process_model}
We consider five process models which describe the transmission mechanism of CWD. Note that combining the two different data models in Section \ref{data_model} with the five process models described below, we consider a total of ten different models for CWD. The five process models, which are based on deterministic or stochastic models, are introduced below.

\subsubsection{CWD direct transmission model}\label{CWD_direct_section}
CWD may be transmitted to susceptible animals directly from infected animals \citep{miller2003prion,miller2006dynamics}. We portray this direct transmission using ODE, CTMC and SDE models.

\paragraph{ODE model}
\citet{miller2006dynamics} propose an ODE model for the direct (animal to animal) transmission of CWD given by
\begin{equation}\label{CWD_direct_ODE}
d\begin{pmatrix}
S\\
I\\
C
\end{pmatrix}=\begin{pmatrix}
a-S(\beta I+m)\\
\beta SI-I(\mu+m)\\
\mu I
\end{pmatrix}dt,
\end{equation}
where $a$ is the known number of susceptible animals annually added to the population via births or importation, $m$ is the known per capita natural mortality rate, $\beta$ is the unknown direct transmission coefficient (unit $=$ time$^{-1}$) and $\mu$ is the unknown per capita CWD mortality rate (unit $=$ time$^{-1}$). The unknown quantities to be estimated are $(\beta,\mu, S(t_0), I(t_0))$, where $S(t_0)$ and $I(t_0)$ are the unknown initial conditions. 

We note that the ODE model \eqref{CWD_direct_ODE} has error variance equal to 0. This is a slight variation on the typical set-up where the process model includes a non-zero error variance \citep[see, e.g.,][Equation (2)]{hooten2008hierarchical}.  It would be possible to incorporate such a model within the framework described in this paper but we did not consider this further here for simplicity.  

\paragraph{CTMC model}
A continuous time Markov chain model can also be  used to study a stochastic epidemic process. In a CTMC model time is continuous, but the random variables of interest are discrete. Based on the direct transmission ODE model \eqref{CWD_direct_ODE}, the probability equations for the CTMC model for the direct transmission of CWD are given by
\begin{subequations}
\label{CTMC}
\begin{align}
P\left(\begin{array}{l|r} \label{CWD_direct_CTMC4}
S(t+\delta)=i+1& S(t)=i\\
I(t+\delta)=j & I(t)=j\\
C(t+\delta)=k &C(t)=k
\end{array}\right)&=a\delta+o(\delta),\\
P\left(\begin{array}{l|r} \label{CWD_direct_CTMC5}
S(t+\delta)=i-1& S(t)=i\\
I(t+\delta)=j & I(t)=j\\
C(t+\delta)=k &C(t)=k
\end{array}\right)&=im\delta+o(\delta),\\
P\left(\begin{array}{l|r} \label{CWD_direct_CTMC6}
S(t+\delta)=i-1& S(t)=i\\
I(t+\delta)=j+1 & I(t)=j\\
C(t+\delta)=k & C(t)=k
\end{array}\right)&=\beta ij\delta+o(\delta),\\
P\left(\begin{array}{l|r} \label{CWD_direct_CTMC7}
S(t+\delta)=i& S(t)=i\\
I(t+\delta)=j-1 & I(t)=j\\
C(t+\delta)=k &C(t)=k
\end{array}\right)&=jm\delta+o(\delta),\\
P\left(\begin{array}{l|r} \label{CWD_direct_CTMC8}
S(t+\delta)=i& S(t)=i\\
I(t+\delta)=j-1 & I(t)=j\\
C(t+\delta)=k+1 &C(t)=k
\end{array}\right)&=j\mu\delta+o(\delta),
\end{align}
 \end{subequations}
where $o(\delta)\rightarrow0$ as the time interval $\delta\rightarrow0$.  Each probability statement in the CTMC model corresponds to a component of the ODE model \eqref{CWD_direct_ODE}.  For example, \eqref{CWD_direct_CTMC4}  is the probability that an additional susceptible deer is added due to birth or importation, \eqref{CWD_direct_CTMC5} accounts for the loss of a susceptible deer due to natural mortality, and \eqref{CWD_direct_CTMC7}  is the corresponding probability for a loss of an infected deer due to natural mortality.  More details about the derivation of a CTMC model based on an ODE model are given by \citet{allen2008introduction}.

\paragraph{SDE model}\label{CWD_direct_SDE_section}
SDE models are a natural extension of ODE models and they may be simpler to derive and apply than Markov chain models. For example, the transition matrix in a continuous time Markov chain model can be very complicated when there are several interacting populations \citep{allen2003comparison, allen2005comparison}. We consider the SDE model for the direct transmission of CWD given by
\begin{equation}\label{CWD_direct_SDE}
d\begin{pmatrix}
S\\
I\\
C
\end{pmatrix}=\begin{pmatrix}
a-S(\beta I+m)\\
\beta SI-I(\mu+m)\\
\mu I
\end{pmatrix}dt+\boldsymbol{B}d\boldsymbol{W},
\end{equation}
where $\boldsymbol{W}=(W_1,W_2,W_3)^T$ is a 3-dimensional standard Wiener process and $\boldsymbol{B}=\sqrt{\boldsymbol{\Sigma}}$ is the positive definite square root of the covariance matrix with
\begin{equation*}\label{Sigma_direct}
\boldsymbol{\Sigma}=\begin{bmatrix}
a+S(\beta I+m) & -\beta SI & 0\\
-\beta SI & \beta SI+I(\mu+m) & -\mu I\\
0 & -\mu I & \mu I
\end{bmatrix}.
\end{equation*}
The derivation of the direct transmission SDE model \eqref{CWD_direct_SDE} is given in \citet{sun2015penalized}; in the next section, we briefly illustrate the derivation of a more complex SDE model for CWD.

\subsubsection{CWD indirect transmission model}\label{CWD_indirect_section}
CWD may also be transmitted to susceptible animals from excreta left in the environment by infected animals. We describe this indirect transmission using  both an ODE and an SDE model. The CTMC model is not suitable here, because excreta left in the environment is not a discrete variable. Let $E$ denote the mass of infectious material in the environment.

\paragraph{ODE model}
An ODE model for the indirect transmission of CWD \citep{miller2006dynamics} is
\begin{equation}\label{CWD_indirect_ODE}
d\begin{pmatrix}
S\\
I\\
E\\
C
\end{pmatrix}=\begin{pmatrix}
a-S(\gamma E+m)\\
\gamma SE-I(\mu+m)\\
\epsilon I-\tau E\\
\mu I
\end{pmatrix}dt,
\end{equation}
where $\gamma$ is the indirect transmission coefficient (unit $=$ mass$^{-1}$time$^{-1}$), $\epsilon$ is the per capita rate of excretion of infectious material by infected animals (unit $=$ time$^{-1}$), and $\tau$ is the mass-specific rate of loss of infectious material from the environment (unit $=$ time$^{-1}$). The unknown quantities to be estimated are $(\gamma,\mu, \epsilon,\tau, S(t_0), I(t_0), E(t_0))$.

\paragraph{SDE model}
The corresponding SDE model for the indirect transmission of CWD can be derived as follows. Let $\boldsymbol{X}(t)$ denote $(S(t),I(t),E(t),C(t))^T$ and $\boldsymbol{X}_\delta=\boldsymbol{X}(t+\delta)-\boldsymbol{X}(t)$ be the increment during the time interval of length of length $\delta$. If $\delta$ is sufficiently small, we can assume at most one animal is added, infected, or died during the time interval of length $\delta$. The probability that more than one addition, infection, or death has occurred during that interval is of order $\delta^2$, which can be neglected. Then we can approximate the mean of $\boldsymbol{X}_\delta$ for sufficiently small $\delta$ by
\begin{equation}\label{indirect_expectation}
E[\boldsymbol{X}_\delta]\approx
\begin{pmatrix}
a-S(\gamma E+m)\\
\gamma SE-I(\mu+m)\\
\epsilon I-\tau E\\
\mu I
\end{pmatrix}\delta \equiv \boldsymbol{f}\delta.
\end{equation}
Furthermore, we can also approximate the covariance of $\boldsymbol{X}_\delta$ for sufficiently small $\delta$ by
\begin{equation}\label{indirect_variance}
\begin{split}
V[\boldsymbol{X}_\delta] &= E[(\boldsymbol{X}_\delta)(\boldsymbol{X}_\delta)^T]-E(\boldsymbol{X}_\delta)E(\boldsymbol{X}_\delta)^T\approx E[(\boldsymbol{X}_\delta)(\boldsymbol{X}_\delta)^T]=\boldsymbol{\Sigma}\delta,
\end{split}
\end{equation}
where $\boldsymbol{\Sigma}$ is the covariance matrix given by
\begin{equation}\label{Sigma_indirect}
\boldsymbol{\Sigma}=\begin{bmatrix}
a+S(\gamma E+m) & -\gamma SE & 0 & 0\\
-\gamma SE & \gamma SE+I(\mu+m) & 0 & -\mu I\\
0 & 0 & \epsilon I+\tau E & 0\\
0 & -\mu I & 0 & \mu I
\end{bmatrix}.
\end{equation}

The matrix $\boldsymbol{\Sigma}$ in \eqref{Sigma_indirect} is positive definite and hence has a positive definite square root $\boldsymbol{B}=\sqrt{\boldsymbol{\Sigma}}$. It can be shown that \eqref{indirect_expectation} and \eqref{indirect_variance} are quantities of order $\delta$. We also assume $\boldsymbol{X}_\delta$ follows a normal distribution with mean vector $\boldsymbol{f}\delta$ and covariance matrix $\boldsymbol{B}^2\delta=\boldsymbol{\Sigma}\delta$. Thus,
\begin{equation}\label{direct_explanation}
\boldsymbol{X}(t+\delta)\approx\boldsymbol{X}(t)+\boldsymbol{f}\delta+\boldsymbol{B}\sqrt{\delta}\boldsymbol{\eta},
\end{equation}
where $\boldsymbol{\eta}\sim N(\boldsymbol{0}, \boldsymbol{\mathcal{I}}_{4\times 4})$ and $\boldsymbol{\mathcal{I}}$ is the identity matrix. This is exactly one iteration of the Euler-Maruyama scheme \citep{kloeden1992numerical} for a system of SDE, which is given by
\begin{equation}\label{CWD_indirect_SDE}
d\begin{pmatrix}
S\\
I\\
E\\
C
\end{pmatrix}=\begin{pmatrix}
a-S(\gamma E+m)\\
\gamma SE-I(\mu+m)\\
\epsilon I-\tau E\\
\mu I
\end{pmatrix}dt+\boldsymbol{B}d\boldsymbol{W},
\end{equation}
where $\boldsymbol{W}=(W_1,W_2,W_3, W_4)^T$ is a 4-dimensional standard Wiener process. The dynamical system \eqref{direct_explanation} converges in the mean square sense to the system of SDEs \eqref{CWD_indirect_SDE} as $\delta\rightarrow 0$ \citep{kloeden1992numerical}.

\subsection{Parameter model}
We consider three sets of prior distributions. The two sets of informative prior distributions which were chosen based on expert knowledge. We selected distributions and elicited distribution moments with the assistance of N. Thompson Hobbs, an expert on CWD.  The parameters  $\beta$, $\mu$, and $\epsilon$ are most likely be between 0 and 1; thus we used a Beta or uniform distribution as the informative priors for these parameters.  Little is known about $\gamma$ and $\tau$ and thus we used less informative prior distributions for these parameters. To investigate sensitivity to these priors, we also consider a set of noninformative prior distributions. The three sets of prior distributions for parameters $\boldsymbol{\theta}$ and initial conditions $(S(t_0),I(t_0),E(t_0))$ are shown in Table \ref{priors}.  In a non-Bayesian context, the parameter model can be omitted. 

\begin{table}[!h]
\caption{The prior distributions for parameters and initial conditions. Recall $\beta$ is the direct transmission coefficient, $\mu$ is the per capita CWD mortality rate, $\gamma$ is the indirect transmission coefficient, $\epsilon$ is the per capita rate of excretion of infectious material by infected animals, and $\tau$ is the mass-specific rate of loss of infectious material from the environment. }
\centering
\label{priors}
\begin{tabular}{ c | c | c | c || c | c}
 & Informative I & Informative II & Noninformative & Initial & Prior\\\hline
$\beta$ & Beta(2,10)  & U(0,1)& Gamma(0.1,0.1)& $S(t_0)$ & Discrete U(10,50)\\
$\mu$ & Beta(2,5) & U(0,1)& Gamma(0.1,0.1) & $I(t_0)$  & Discrete U(0,20)\\
$\gamma$ & Gamma(0.01,0.01) &U(0,20) &Gamma(0.1,0.1)& $E(t_0)$ & U(0,6)\\
$\epsilon$ & Beta(2,2)& U(0,1) &Gamma(0.1,0.1) \\
$\tau$ & Gamma(0.01,0.01) & U(0,20)&Gamma(0.1,0.1)\\
\end{tabular}
\end{table}

\section{Approximate Bayesian computation (ABC)}\label{ABC}
For all the process models described in Sections \ref{CWD_direct_section} and \ref{CWD_indirect_section}, we assume the data model is given in \eqref{data_binom} or \eqref{data_pois}. That is, only $\tilde{C}(t)$ is observed at discrete time $t=t_0,t_1,\dots,t_n$. To estimate the parameters in the process models via maximum likelihood, one needs to compute the likelihood,
\begin{equation}\label{likelihood}
\int\cdots\int \prod_{i=0}^n\left[p\left(\tilde{C}(t_i)|\boldsymbol{X}(t_i),\boldsymbol{\theta}\right)p\Big(\boldsymbol{X}(t_{i+1})|\boldsymbol{X}(t_i),\boldsymbol{\theta}\Big)\right]d\boldsymbol{X}(t_0)\cdots d\boldsymbol{X}(t_n),
\end{equation}
where $p(\tilde{C}(t_i)|\boldsymbol{X}(t_i),\boldsymbol{\theta})$ is given by \eqref{data_binom} or \eqref{data_pois} and $\boldsymbol{X}(t)\equiv\left(S(t), I(t), C(t)\right)^T$ or $(S(t), I(t), E(t), C(t))^T$, depending on the process model that is assumed. The likelihood \eqref{likelihood} thus requires a multivariate integration over all unobserved state variables $\boldsymbol{X}(t)$, which can be computationally intensive or even infeasible.

To carry out Bayesian inference using a Markov chain Monte Carlo algorithm, one can treat all unobserved state variables $\boldsymbol{X}(t)$ as augmented data to avoid this complex integration \citep{golightly2005bayesian, golightly2006bayesian, golightly2008bayesian}. However, MCMC approaches are typically slow to converge for nonlinear multivariate dynamical models, particularly when the time interval between consecutive observations is large \citep{golightly2008bayesian, donnet2011algorithm}, which is often the situation for ecological or environmental data. For example, in the CWD epidemic the number of deaths were recorded annually. In contrast to the slow convergence in MCMC approaches, simulating data from the process models is relatively straightforward. For example, there are many numerical methods for solving ODEs, such as Euler's method and the Runge-Kutta method \citep{butcher2008numerical}. Based on the Markov property, simulating sample paths of a CTMC is straightforward \citep[Chapter~5]{allen2003introduction}. Simple numerical solutions for SDEs include the Euler-Maruyama and the Milstein methods \citep{kloeden1992numerical}. Embedding these simulation methods in the approximate Bayesian computation with sequential Monte Carlo algorithm makes it a suitable choice for parameter inference and model selection for hierarchical models that are built upon dynamical processes.

The basic idea of ABC is that sample parameters are proposed from their corresponding prior distributions and data are simulated from the model based on the proposed parameters. The proposed parameters are accepted if the difference between the summary statistics $\eta(\cdot)$ of the simulated data $D^*$ and the observed data $D$ is small. The simplest ABC approach is the ABC rejection algorithm proposed by \citet{tavare1997inferring} and \citet{pritchard1999population}. In the ABC SMC algorithm \citep{toni2009approximate}, $N$ samples of parameters $\boldsymbol{\theta}$ are proposed through a sequence of intermediate distributions, $f(\boldsymbol{\theta}|\rho(\eta(D^*),\eta(D))\leq \xi_t)$, with decreasing distance tolerances, $\xi_1>\dots>\xi_T>0$, between prior distribution and target distribution, $f(\boldsymbol{\theta}|\rho(D^*,D)\leq \xi_T)$. Here, $\rho$ is a distance function between the summary statistics $\eta(\cdot)$ of the simulated data $D^*$ and the observed data $D$. For each distance tolerance $\xi_t$, $1\leq t\leq T$, a new candidate sample parameter $\boldsymbol{\theta}^{**}$ is drawn from a proposal distribution $q_t(\boldsymbol{\theta}|\boldsymbol{\theta}^*)$, where $\boldsymbol{\theta}^*$ is a sample from the previous population of all proposals that have a distance tolerance $\xi_{t-1}$. The advantage of generating samples via a sequence of distributions is that it often avoids the problem of having low acceptance rates which is common in ABC rejection and ABC MCMC algorithms \citep{toni2009approximate}. The ABC SMC algorithm is given in Algorithm \ref{alg} (\citet{toni2009approximate} provide a similar algorithm).

\RestyleAlgo{boxruled}
\begin{algorithm}[ht]
\caption{The ABC SMC algorithm.\label{alg}}
\noindent\fcolorbox{white}{white}{\parbox{0.95\textwidth}{
\begin{enumerate}[Step 1.]
\item Set the tolerance sequence $\xi_1>\dots>\xi_T>0$, and $t=1$.
\item Set the sample index $i=1$.
\item Sample model index $\mathcal{M}^*$ from the model prior $\pi(\mathcal{M})$. If $t=1$, sample $\boldsymbol{\theta}^{**}$ from the prior distribution $\pi_{\mathcal{M}^*}(\boldsymbol{\theta})$. Else, sample $\boldsymbol{\theta}^{*}$ from the previous population $\{\boldsymbol{\theta}_{t-1,\mathcal{M}^*}^{(i)}\}$ with weights $\omega_{t-1,\mathcal{M}^*}$ and sample $\boldsymbol{\theta}^{**}$ from the proposal distribution $q_t(\boldsymbol{\theta}|\boldsymbol{\theta}^*)$.
\item If $\pi_{\mathcal{M}^*}(\boldsymbol{\theta}^{**})=0$, return to Step 3.
\item Simulate $B_t$ candidate datasets, $D_1, D_2, \dots, D_{B_t}$, based on candidate parameter $\boldsymbol{\theta}^{**}$ and model $\mathcal{M}^*$. Calculate $b_t(\boldsymbol{\theta}^{**})=\sum_{b=1}^{B_t}I(\rho(D_b, D)\leq \xi_t)$, where $I(x)$ is the indicator function.
\item If $b_t(\boldsymbol{\theta}^{**})=0$, return to Step 3.
\item Update $\mathcal{M}_t^{(i)}=\mathcal{M}^*$ and $\boldsymbol{\theta}_t^{(i)}(\mathcal{M}^*)=\boldsymbol{\theta}^{**}$. Update its weight,
\begin{equation*}
\omega_{t,\mathcal{M}^*}^{(i)}=\left\{\begin{array}{ll}
b_t(\boldsymbol{\theta}^{**}), & \text{if } t=1,\\
\frac{\pi_{\mathcal{M}^*}(\boldsymbol{\theta}^{**})b_t(\boldsymbol{\theta}^{**})}{\sum_{j=1}^{N_{\mathcal{M}^*}}\omega_{t-1,\mathcal{M}^*}^{(j)}q_t\left(\boldsymbol{\theta}^{**}|\boldsymbol{\theta}_{t-1,\mathcal{M}^*}^{(j)}\right)}, & \text{if } t>1,
\end{array}\right.
\end{equation*}
where $N_{\mathcal{M}^*}$ is the number of samples for the model $\mathcal{M}^*$.
\item If $i<N$, update $i=i+1$ and go to Step 3.
\item Normalize the weights for each model $\mathcal{M}$. If $t<T$, update $t=t+1$ and go to Step 2.
\end{enumerate}
}}
\end{algorithm}

Step 3 of Algorithm \ref{alg} requires selection of a proposal distribution from which to sample a set of candidate parameters.  We chose the  proposal distribution $q_t(\boldsymbol{\theta}|\boldsymbol{\theta}^*)$ to be a normal or uniform random walk (that is, $\theta=\theta^*+\zeta$, where $\zeta$ is sampled from a normal or uniform distribution).  We discuss this further for the specific examples below and in Table~\ref{proposal}.  

ABC also requires selection of a number of parameters and functions including selection of a set of summary statistics $\eta$, a distance function $\rho$, and two tuning parameters, $\boldsymbol{\xi}=\{\xi_1,\dots,\xi_T\}$ and $B_t$.    The determination of summary statistics requires some care.  \cite{marin2014} showed that model selection via ABC  is only consistent when the summary statistics are either the full dataset or a set of sufficient statistics that are sufficient under all models under consideration (see Section 2.1 of Marin et al. for additional discussion of these requirements).  For our problem involving  discrete-time observations of a dynamical process, no summary statistics are required because we can compare the simulated and observed datasets directly, so $\eta(D)=D$.  
In general one reasonable choice of the distance function $\rho(D^*, D)$ is $\frac{1}{n}\sum_i\lVert\mathbf{x}^*_i-\mathbf{x}_i\rVert$, where $\mathbf{x}^*_i$ and $\mathbf{x}_i$ are the corresponding $i$th observation in the simulated dataset $D^*$ and observed dataset $D$, respectively, and $\lVert\cdot\rVert$ is a Euclidean norm for this case. A similar distance function is used in \citet{toni2009approximate}. For the one dimensional CWD cumulative death data, this reduces to  $\rho(D^*, D)=\frac{1}{n}\sum_i|\tilde{C}^*(t_i)-\tilde{C}(t_i)|$; hence it is equivalent to use the $L^1$ or infinity norm. The vector $\boldsymbol{\xi}$ such that $\xi_1>\dots>\xi_T>0$ denotes the tolerance level for the cut-off for the distance function, $\rho(D^*,D)\leq\xi_i$ for $i=1,\dots,T$. Note that the tolerance level $\boldsymbol{\xi}$ does not have a strong influence on ABC output, but computational costs are significantly  increased as $\boldsymbol{\xi}$ decreases \citep{marin2012approximate}. In practice one can select $\boldsymbol{\xi}$ as a small percentile of the simulated distance $\rho(D^*, D)$  \citep{beaumont2002approximate}. $B_t$ is the number of simulated datasets for a given parameter $\boldsymbol{\theta}$ for stochastic models. For the deterministic model, one uses $B_t=1$. A larger $B_t$ may decrease the computational time of  the ABC algorithm because it allows the algorithm more opportunities to generate a dataset that is sufficiently close to the observed dataset. For our model set-up we have found that  using $B_t=5$ or $10$ is generally sufficient.

The outputs of the ABC SMC algorithm are the approximations of the marginal posterior distribution of the model parameter $P(\mathcal{M}|D)$, which is the proportion of times that model $\mathcal{M}$ is selected in $N$ samples, and the marginal posterior distributions of parameters $P(\boldsymbol{\theta}|D,\mathcal{M})$ for models $\mathcal{M}=\mathcal{M}_1,\mathcal{M}_2,\dots,\mathcal{M}_M$. We consider the ABC SMC algorithm in a model selection context where we simultaneously estimate parameters and perform model selection. 

Consider the problem where one wishes to compare the posterior distributions of two models, $P(\mathcal{M}_1|D)$ and $P(\mathcal{M}_2|D)$. The ABC SMC output can be used to perform model selection based on the Bayes factor,
\begin{equation}\label{Bayes_factor}
B_{\mathcal{M}_1\mathcal{M}_2}=\frac{P(\mathcal{M}_1|D)/P(\mathcal{M}_2|D)}{\pi(\mathcal{M}_1)/\pi(\mathcal{M}_2)},
\end{equation}
where $\pi(\mathcal{M})$ is the model prior. A commonly used interpretation of the Bayes factor values, which is  given by \citet{kass1995bayes}, is shown in Table \ref{Bayes}. In this work we adopt the model prior $\pi(\mathcal{M})$ as the discrete uniform distribution from $1$ to $M$ for models $\mathcal{M}_1$ to $\mathcal{M}_M$.  

\begin{table}[h]
\centering
\caption{Interpretation of the Bayes factor, where ``strength of evidence" indicates evidence in favor of model 1 against model 2.}
\label{Bayes}
\begin{tabular}{ c|c }
The Bayes factor $B_{12}$ & Strength of evidence\\\hline
1 to 3 & Weak\\
3 to 20 & Positive \\
20 to 150 & Strong\\
$>$150 & Very Strong
\end{tabular}
\end{table}

\section{Simulation studies}\label{simulation}
We illustrate the performance of the ABC SMC algorithm on 100 simulated datasets. Each dataset includes 21 annual CWD death observations from two distinct CWD epidemics similar to the observed epidemic data in Section \ref{app results}. We generate 100 datasets under two different scenarios: (a) the indirect transmission SDE process model \eqref{CWD_indirect_SDE} with the Binomial data model \eqref{data_binom}, parameters $(\gamma_0,\mu_0,\epsilon_0,\tau_0)=(0.15, 0.20, 0.50, 1.70)$, and a set of initial conditions for each epidemic given by $(S(t_0), I(t_0), E(t_0))=(24,5,4.04)$ and $(22,2,0.87)$; (b) the direct transmission CTMC process model \eqref{CTMC} with the Binomial data model \eqref{data_binom},  parameters $(\beta_0,\mu_0)=(0.04,0.30)$, and initial conditions $(S(t_0), I(t_0))=(12,14)$ and $(30,5)$. The parameters and initial conditions were selected so that the simulated trajectories are similar to the observed data (Section \ref{app results}). We apply the ABC SMC algorithm on each dataset for parameter estimation and model selection among the ten models (five process models and two data models) described in Section \ref{models}. The set-up for the ABC SMC algorithm is the same as the set-up we used for the observed real data and is described in Section \ref{app results}.

To investigate model selection performance of the ABC SMC algorithm, we record the number of times that the true model (the indirect transmission SDE process model \eqref{CWD_indirect_SDE} with the Binomial data model \eqref{data_binom} or the direct transmission CTMC process model \eqref{CTMC}  with the Binomial data model \eqref{data_binom}) has the highest posterior model probability $P(\mathcal{M}|D)$ among the ten models for the 100 simulated datasets.  We compute the Bayes factor between the true model and the model that has the highest probability for 100 simulated datasets for two scenarios.

\begin{figure}[!h]
\centerline{\includegraphics[width=0.9\textwidth]{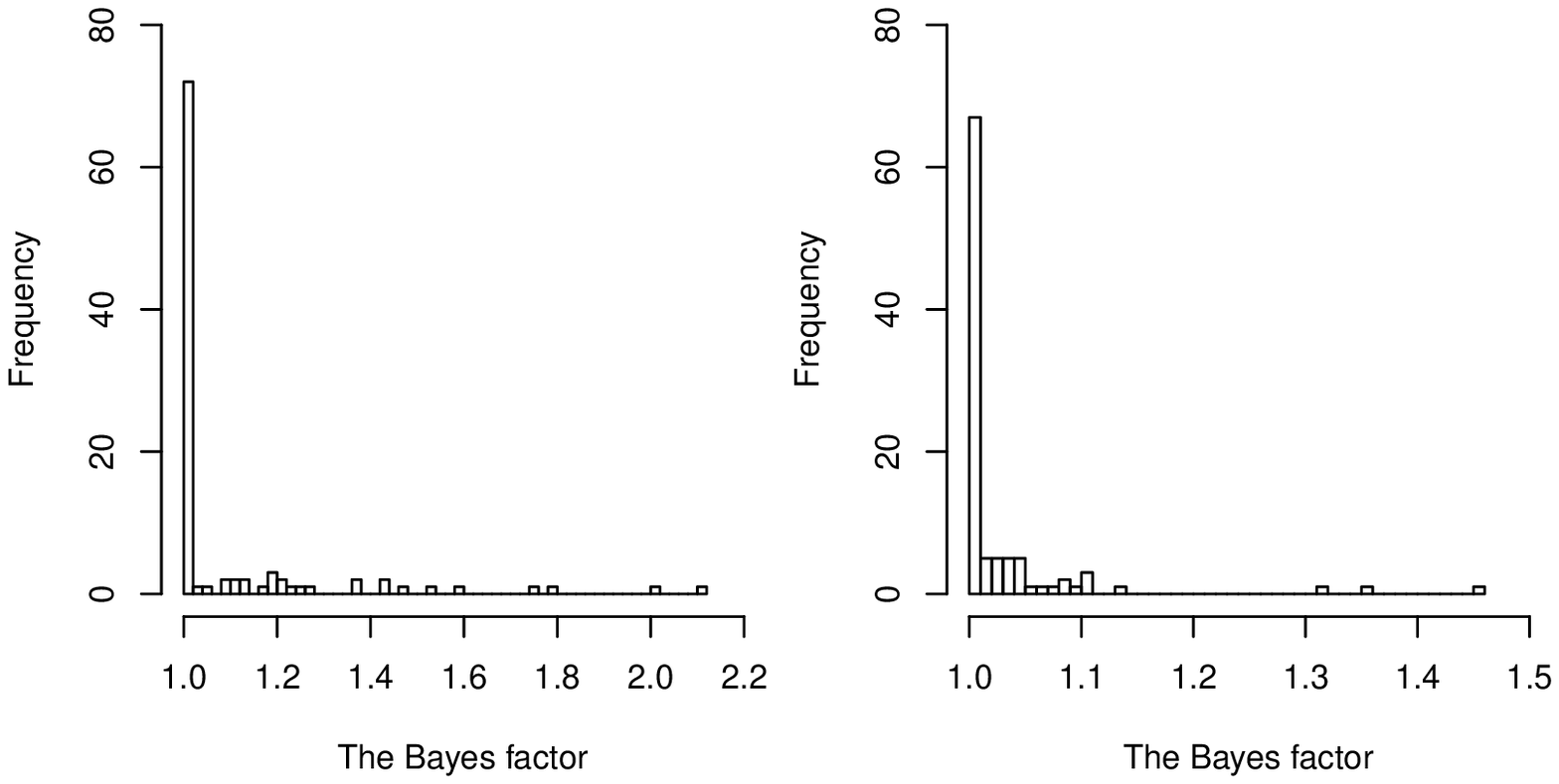}}
\vspace{-2.8in}
\caption{The histogram of the Bayes factor in favor of the model with the highest posterior model probability against the true model for 100 simulated datasets (scenario (a) left: the indirect transmission SDE model \eqref{CWD_indirect_SDE} with the Binomial data model \eqref{data_binom}; scenario (b) right: the direct transmission CTMC process model \eqref{CTMC}  with the Binomial data model \eqref{data_binom}). Note that if the true model has the highest posterior model probability then the Bayes factor is 1.}
\label{simulation_bayes}
\end{figure}

For the first scenario, in 71 out of the 100 simulated datasets the true model (the indirect transmission SDE process model \eqref{CWD_indirect_SDE} with the Binomial data model \eqref{data_binom}) has the highest posterior model probability among the ten models. Figure \ref{simulation_bayes} left shows the histogram of the Bayes factor in favor of the model with highest posterior model probability against the true model over the 100 simulated datasets. Note that if the true model has the highest posterior model probability then the Bayes factor is 1. In 91 out of the 100 simulated datasets, the Bayes factor is less than 1.4. In fact, there is no dataset for which the Bayes factor is larger than 2.2. Although the ABC SMC algorithm does not always select the true model as the highest probability model, it is apparent that the strength of evidence in favor of the other models is very weak. For the second scenario, similar results are obtained. The true model (the direct transmission CTMC process model \eqref{CTMC}  with the Binomial data model \eqref{data_binom}) was selected as the best model for 64 out of 100 simulated datasets (Figure \ref{simulation_bayes} right) and as the second best model for 28 simulated datasets. The closest model to the true model, the direct transmission CTMC process model \eqref{CTMC} with the Poisson data model \eqref{data_pois}, is selected as the best model in 25 simulated datasets and as second best model in 50 simulated datasets. 

Note that in the ABC algorithm the proposed parameter is accepted if the simulated data based on it are close enough to the observed data. If the observed data were generated from a bad model, then the simulated data from the candidate model probably will be far away from the observed data. Hence, no proposed parameter will be accepted and the ABC algorithm will be unlikely to converge. 

\section{CWD application results}\label{app results}
We apply the ABC SMC algorithm to the CWD epidemic data, which includes 21 annual CWD death observations from two distinct CWD epidemics as described in Section \ref{CWD}.  To carry out model selection we compute the posterior model probability $P(\mathcal{M}|D)$ for each model and the Bayes factors to compare pairs of models. We compare the ten models in Section \ref{models} and  assume all models are equally likely by adopting a discrete uniform distribution as the prior distribution of the model parameter $\mathcal{M}$. We consider three sets of prior distributions for the other model parameters (Table \ref{priors}). 

For the ABC SMC algorithm the tolerance sequence is set to be $\xi=\{7,6,5,4,3.5,3\}$, so $T=6$, and $N=2500$ samples of parameters are generated. The proposal distributions $q_t$ for parameters $\boldsymbol{\theta}$ and initial conditions $(S(t_0),I(t_0),E(t_0))$ are based on a random walk described in Table \ref{proposal}.  For example, $\beta^{(t)}\sim N(\beta^{(t-1)},0.02^2)$.   We chose a small variance for the proposal distribution for the parameters $\beta$, $\mu$, $\gamma$ and $\epsilon$ because these parameters are generally small.  The parameter $\tau$ takes on larger values so we use a larger variance. The simulated data from  the ODE models, \eqref{CWD_direct_ODE} and \eqref{CWD_indirect_ODE}, are generated using the \texttt{ode} function with default settings in the \texttt{deSolve} R package \citep{desolve}. The simulation method described in \citet[Chapter~5]{allen2003introduction} is used for simulating  the CTMC process model \eqref{CTMC}. The sample paths of the SDE models, \eqref{CWD_direct_SDE} and \eqref{CWD_indirect_SDE}, are approximated using the Euler-Maruyama scheme \citep{kloeden1992numerical} with time step $\delta=1/12$ which is one month for the CWD epidemic data.

\begin{table}[!h]
\caption{The proposal distributions for model parameters. We used a random walk proposal for each parameter. Below, the superscripts $(t)$ and $(t-1)$ refer to iteration number in the ABC SMC algorithm. The initial conditions $S(t_0)$,  $I(t_0)$ and $E(t_0)$ are the unknown values of the number of susceptible and infected animals, the unknown mass of infectious material in the environment at time $t_0$, respectively.}
\centering
\label{proposal}
\begin{tabular}{ l || l}
Parameter \& proposal distribution $q_t$ & Initial condition parameters \& proposal distribution $q_t$\\\hline
$\beta^{(t)} \sim N(\beta^{(t-1)},0.02^2)$ & $S(t_0)^{(t)} = S(t_0)^{(t-1)}+\zeta$ where $\zeta \sim\mbox{Discrete U}(-8,8)$\\
$\mu ^{(t)}\sim N(\mu ^{(t-1)},0.2^2)$ &  $I(t_0)^{(t)} = I(t_0)^{(t-1)}+\zeta$ where $\zeta\sim\mbox{Discrete U}(-3,3)$\\
$\gamma ^{(t)}\sim N(\gamma ^{(t-1)},0.2^2)$ & $E(t_0)^{(t)} = E(t_0)^{(t-1)}+\zeta$ where $\zeta \sim \mbox{U}(-1,1)$\\
$\epsilon^{(t)}\sim N(\epsilon^{(t-1)},0.2^2)$ \\
$\tau^{(t)}\sim N(\tau^{(t-1)},4)$
\end{tabular}
\end{table}

Posterior model probabilities $P(\mathcal{M}|D)$ and the Bayes factor in favor of the model constructed with the indirect SDE process model \eqref{CWD_indirect_SDE} with the Binomial data model \eqref{data_binom} against the other models are shown in Table \ref{m.model.table}. The Bayes factor results indicate that the Binomial data model \eqref{data_binom} is generally preferred over the Poisson data model \eqref{data_pois}. The marginal posterior model probability of the Binomial data model \eqref{data_binom} practically remains unchanged under different prior sets, 0.72, 0.73, and 0.72. There is uncertainty about the form of the process model.  There is a weak evidence in favor of the indirect transmission SDE process model \eqref{CWD_indirect_SDE} compared with the other process models considered in Section \ref{Process_model}.  It is of particular interest to biologists about whether the indirect CWD transmission model is supported by the data because indirect transmission makes CWD control efforts very challenging \citep{miller2006dynamics}. For the informative prior set I there are no significant differences among the other four process models in terms of the Bayes factor. Since the evidence in favor of the indirect transmission SDE model \eqref{CWD_indirect_SDE} is not very strong, one could consider Bayesian model averaging  \citep{hoeting1999bayesian}. Model averaging can provide more accurate forecasts if the goal is to predict the development of the disease in the future. The results based on different prior sets are similar. It appears that the ABC SMC model selection is not sensitive to the priors we used for this study. The main difference is in the ordering the direct versus the indirect ODE process model under the binomial data model.  

\begin{table}[!h]
\caption{Posterior model probabilities for each model $P(\mathcal{M}|D)$ and the Bayes factor (BF) in favor of the indirect SDE process model \eqref{CWD_indirect_SDE} with the Binomial data model \eqref{data_binom} against the other models for the CWD epidemic data based on 2500 samples of parameters of ABC SMC.  The results are given in order of the posterior model probabilities $P(\mathcal{M}|D) $ from the informative prior set I. The three prior sets are listed in Table \ref{priors}. }
\centering
\label{m.model.table}
\begin{tabular}{ ll | cc | cc | cc}
Data & Process&\multicolumn{2}{c|}{Informative I} & \multicolumn{2}{c|}{Informative II} & \multicolumn{2}{c}{Noniformative}\\
Model & Model & $ P(\mathcal{M}|D) $ & BF & $P(\mathcal{M}|D) $ & BF & $P(\mathcal{M}|D) $ & BF \\\hline
Binom (1) & Indirect SDE \eqref{CWD_indirect_SDE} & 0.21 & 1.00 &  0.26 & 1.00 & 0.20 & 1.00\\ 
Binom (1) & Direct SDE \eqref{CWD_direct_SDE} & 0.18 & 1.15 & 0.18 & 1.41 & 0.17 & 1.17 \\
Binom (1) & Direct ODE \eqref{CWD_direct_ODE}  & 0.13 & 1.55 & 0.06 & 3.99 & 0.13 & 1.52\\ 
Binom (1) & Direct CTMC \eqref{CTMC}  & 0.11 & 1.87 &0.08 & 3.20 & 0.11 & 1.89\\ 
Binom (1) & Indirect ODE \eqref{CWD_indirect_ODE}  & 0.09 & 2.43 & 0.15 & 1.71 & 0.11 & 1.83 \\ 
Pois (2) & Indirect SDE \eqref{CWD_indirect_SDE} & 0.09 & 2.27 & 0.08 & 3.30 & 0.06 & 3.48\\
Pois (2) & Direct ODE \eqref{CWD_direct_ODE} & 0.06 & 3.48 & 0.03 & 9.24 & 0.06 & 3.15\\  
Pois (2) & Direct SDE \eqref{CWD_direct_SDE} & 0.05 & 3.87 & 0.06 & 4.20 & 0.08 & 2.64 \\ 
Pois (2) & Indirect ODE \eqref{CWD_indirect_ODE} & 0.04 & 4.63 &  0.07 & 3.92 & 0.04 & 4.60\\
Pois (2) & Direct CTMC \eqref{CTMC}  & 0.03 & 6.17 & 0.03 & 9.66 & 0.04 & 5.06 \\ 
\end{tabular}
\end{table}

\begin{figure}
\centerline{\includegraphics[width=0.9\textwidth]{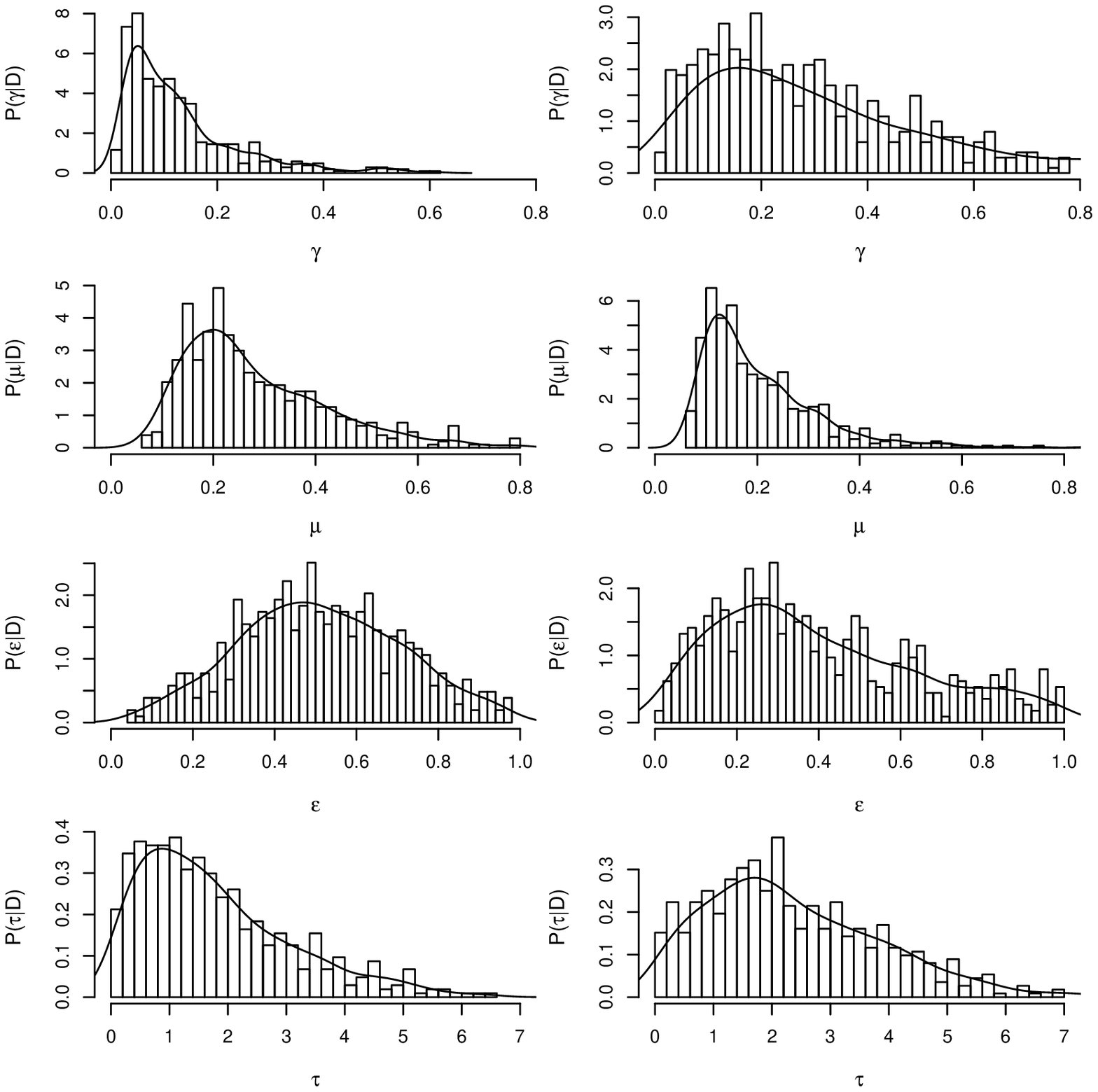}}
\caption{The marginal posterior distribution for the parameters of the indirect transmission SDE process model \eqref{CWD_indirect_SDE} with the Binomial data model \eqref{data_binom} based on the CWD epidemic data. The left column is based on the informative prior set I and the right column is based on the informative prior set II listed in Table \ref{priors}. A smoothed density has been super-imposed.}
\label{posterior_theta}
\end{figure}

The marginal posterior distributions for the parameters for the indirect process model \eqref{CWD_indirect_SDE} and Binomial data model \eqref{data_binom} are given in Figure \ref{posterior_theta} and Table \ref{posteriors} for the two informative prior sets. The results from the noninformative prior set is very similar (not shown). The results under the two sets of informative prior distributions are similar except the parameter $\gamma$ which models the indirect transmission rate. This parameter is particularly challenging to estimate as we are estimating the effects due to some unknown mass of infectious material in the environment (see Section \ref{CWD_indirect_section}). The influence of the prior on the estimates of $\gamma$ and the wide highest posterior density (HPD) intervals are probably due to the small sample size. There is also a considerable uncertainty about $\epsilon$, the per capita rate of excretion of the infectious agent, for both prior sets.  This is not surprising as this transmission mechanism is difficult to quantify.  While it has been demonstrated that CWD can be transmitted via the environment, the scientific community is still trying to understand the exact mechanisms of its transmission. Although the modes of the estimated density of the parameters are different based on the different prior sets, the HPD intervals for $\mu$, $\epsilon$, $\tau$, and initial conditions are similar (Table \ref{posteriors}).    

\begin{table}[!h]
\caption{The marginal posterior modes and $95\%$ highest posterior density (HPD) intervals of the parameters of the indirect transmission SDE process model \eqref{CWD_indirect_SDE} with the Binomial data model \eqref{data_binom} based on the CWD epidemic data. }
\centering
\label{posteriors}
\begin{tabular}{l| l| l| l| l }
\multicolumn{1}{l|}{} & \multicolumn{2}{l|}{Informative prior set I} & \multicolumn{2}{l}{Informative prior set II}\\
Parameter& Mode &95\% HPD & Mode &95\% HPD\\ \hline
$\gamma $ \small{(Indirect transmission rate (mass$^{-1}$yr$^{-1}$))} &0.05&$(0.01, 0.36)$ & 0.16 & (0.02,0.63) \\
 $\mu$ \small{(CWD mortality rate (yr$^{-1}$))}&0.20&$(0.10, 0.59)$ & 0.12 & (0.07,0.41)\\
$\epsilon$ \small{(Per capita rate of excretion of infectious agent (yr$^{-1}$))}&0.47&$(0.15,0.91)$ & 0.26 & (0.02,0.89)\\
$\tau$ \small{(Rate of loss of infectious agent  (yr$^{-1}$))}& 0.88&$(0.01, 4.52)$ & 1.71 & (0.01,5.07)\\ \hline\hline
$S(0)$ of the first epidemic & 18 & (10,26) & 20 & (11,37)\\
$I(0)$ of the first epidemic & 10 & (5,18) & 16 & (0,18)\\
$E(0)$ of the first epidemic & 1.73 & (0.97,5.84) & 4.93 & (0.48,5.94)\\\hline
$S(0)$ of the second epidemic & 48 & (24,50) & 28 & (20,48)\\
$I(0)$ of the second epidemic & 2 & (0,5) & 1 &(0,5)\\
$E(0)$ of the second epidemic & 3.47 & (0.24,4.85) & 1.11 & (0.02,4.59)\\
\end{tabular}
\end{table}

\begin{figure}
 \centerline{\includegraphics[width=0.9\textwidth]{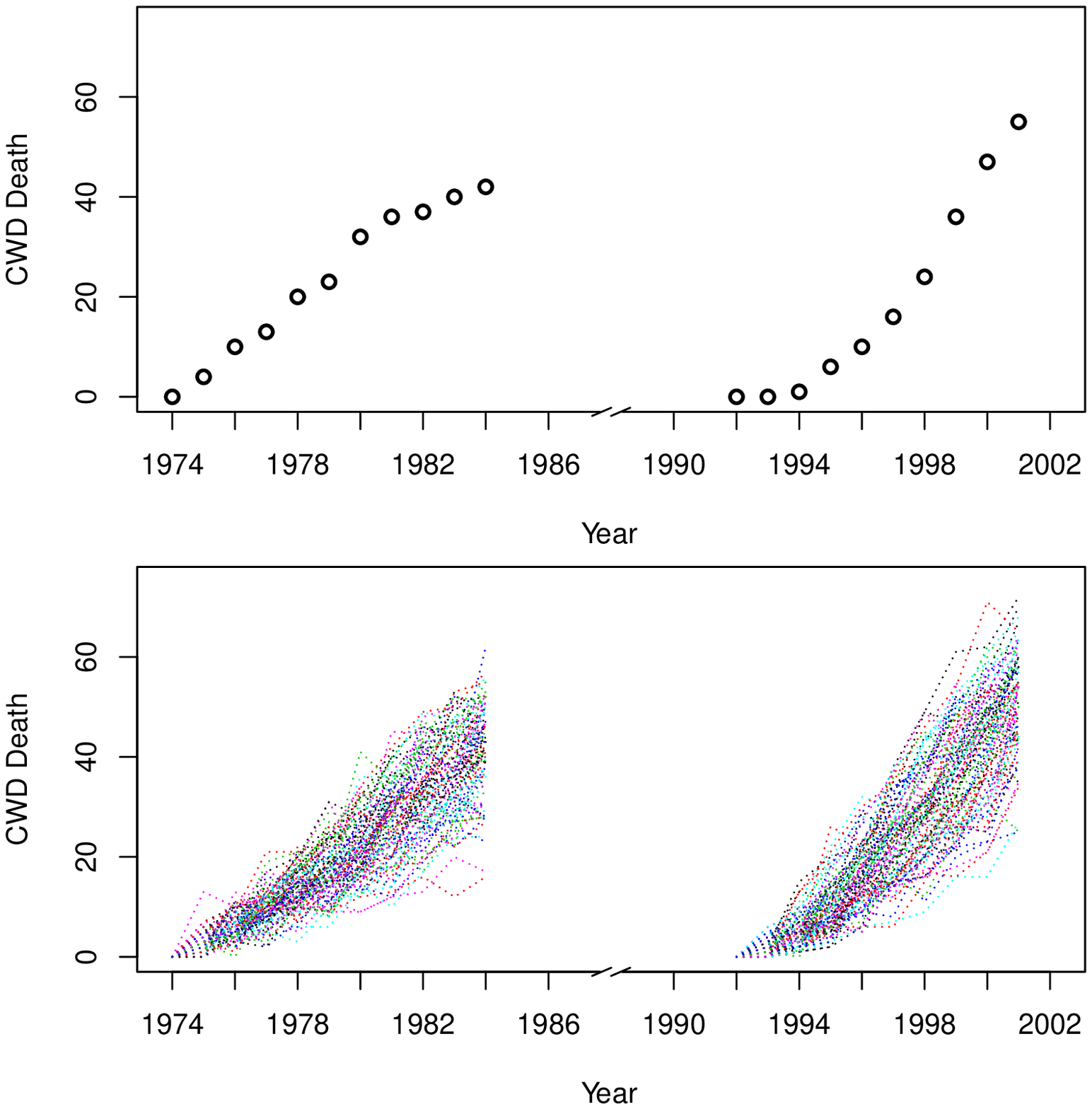}}
\caption{Upper display: the observed cumulative number of deaths for CWD. Lower display: the 100 simulated trajectories of the cumulative number of deaths for CWD are obtained by using the CWD indirect transmission SDE process model \eqref{CWD_indirect_SDE} with the Binomial data model \eqref{data_binom} and posterior estimates of both the parameters and the initial conditions from ABC SMC.}
\label{CWD_indirect_fig}
\end{figure}

To assess goodness of fit, we generated 100 simulated trajectories of the cumulative number of deaths for CWD. To construct the trajectories we used the CWD indirect transmission SDE process model \eqref{CWD_indirect_SDE} with the Binomial data model \eqref{data_binom} and the modes of the estimated density of the parameters from the informative prior set I listed in Table \ref{posteriors}. The simulated trajectories and the observed CWD data are overlaid in Figure \ref{CWD_indirect_fig}. The simulated trajectories based on the mode estimates from the noninformative prior set are very similar (results not shown). If the dataset had more observations we would predict a hold-out set, but this is not reasonable for these data.  The simulated trajectories in Figure \ref{CWD_indirect_fig} are close to the observed data given that they were based on a theoretical model for the process and not from a purely empirical model based only on the observed counts.

\section{Conclusion and discussion}\label{conclusion}
In the pursuit of gaining further understanding of ecological or environmental processes, it is important for statisticians to continue to develop tools for parameter inference and model selection for complex models. The parameters and models for the description of the transmission of CWD play a vital role in its ecological interpretation. A choice between deterministic or stochastic dynamic models is typically based on a scientific theory or personal (ad hoc) preference. We offer a systematic approach to select among these models based on empirical evidence. Although there has been considerable research focused on selecting ecological or environmental models among deterministic models, we are not aware of any previous work where deterministic and stochastic models are directly compared and selected. We illustrate a real world example which considers both deterministic and stochastic models based on the observed data via the ABC SMC algorithm. Simulation studies show the effectiveness of this approach.

We used Bayes factors for model selection because they are easy to calculate using ABC SMC.   As described in Section~\ref{ABC},  some care must be taken to ensure that the model selection results based on ABC are consistent.  This has been an area of recent interest \citep{robert2011,marin2014}.  
There are many other options for model selection in addition to Bayes factors such as the deviance information criteria (DIC) \citep{spiegelhalter2002bayesian}. All commonly used model selection methods have some desirable theoretical properties but there is no single method that can be used for all situations. For example, Bayes factors can be hard to estimate for some models and DIC can give incorrect results when the posterior distribution is not well summarized by the mean \citep{gelman2014understanding}. Most methods can give misleading results if the statistical model is inappropriate \citep[e.g.,][]{hoeting2006model,tenan2014bayesian}. The debate about the properties of different model selection methods will continue  and new model selection methods will continue to be proposed for the foreseeable future  \citep[e.g.,][]{bove2013approximate,watanabe2010asymptotic}.

The choice of distance function or summary statistics used in the ABC SMC algorithm is still an open research topic because  sufficient statistics are not available for many applications. \citet[][Section 2]{marin2014} give guidelines for deciding when a set of statistics is appropriate for ABC. \citet{fearnhead2012constructing} proposed a semi-automatic approach that can construct appropriate summary statistics for ABC. For the CWD epidemic models that we considered here, we found that this approach increases the complexity and decreases the efficiency of the ABC SMC algorithm.

\section*{Acknowledgments}
This material is based upon work supported by the National Science Foundation (EF-0914489, Sun and Hoeting) and the Army Research Office (W911NF-14-1-0216, Lee). This research also utilized the CSU ISTeC Cray HPS System, which is supported by NSF Grant CSN-0923386. We are grateful to N. Thompson Hobbs for fruitful discussions and Michael W. Miller and the Colorado Division of Parks and Wildlife for sharing the data. We appreciate the reviewers for their insightful suggestions that have greatly enhanced the manuscript.

\bibliographystyle{apalike} \bibliography{mybib}

\begin{thebibliography}{}

\bibitem[Allen et~al., 2005]{allen2005comparison}
Allen, E.~J., Allen, L.~J., and Schurz, H. (2005).
\newblock A comparison of persistence-time estimation for discrete and
  continuous stochastic population models that include demographic and
  environmental variability.
\newblock {\em Mathematical Biosciences}, 196(1):14--38.

\bibitem[Allen, 2003]{allen2003introduction}
Allen, L.~J. (2003).
\newblock {\em An Introduction to Stochastic Processes with Applications to
  Biology}.
\newblock Pearson/Prentice Hall Upper Saddle River (New Jersey).

\bibitem[Allen, 2008]{allen2008introduction}
Allen, L.~J. (2008).
\newblock An introduction to stochastic epidemic models.
\newblock In {\em Mathematical epidemiology}, pages 81--130. Springer.

\bibitem[Allen and Allen, 2003]{allen2003comparison}
Allen, L.~J. and Allen, E.~J. (2003).
\newblock A comparison of three different stochastic population models with
  regard to persistence time.
\newblock {\em Theoretical Population Biology}, 64(4):439--449.

\bibitem[Anderson and May, 1992]{anderson1992infectious}
Anderson, R. and May, R. (1992).
\newblock {\em Infectious Diseases of Humans: Dynamics and Control}.
\newblock Wiley Online Library.

\bibitem[Beaumont, 2010]{beaumont2010approximate}
Beaumont, M.~A. (2010).
\newblock Approximate {B}ayesian computation in evolution and ecology.
\newblock {\em Annual Review of Ecology, Evolution, and Systematics},
  41:379--406.

\bibitem[Beaumont et~al., 2002]{beaumont2002approximate}
Beaumont, M.~A., Zhang, W., and Balding, D.~J. (2002).
\newblock Approximate {B}ayesian computation in population genetics.
\newblock {\em Genetics}, 162(4):2025--2035.

\bibitem[Berliner, 1996]{berliner1996hierarchical}
Berliner, L.~M. (1996).
\newblock Hierarchical bayesian time series models.
\newblock In {\em Maximum entropy and Bayesian methods}, pages 15--22.
  Springer.

\bibitem[Bov{\'e} and Held, 2013]{bove2013approximate}
Bov{\'e}, D.~S. and Held, L. (2013).
\newblock Approximate {B}ayesian model selection with the deviance statistic.
\newblock {\em arXiv preprint arXiv:1308.6780}.

\bibitem[Butcher, 2008]{butcher2008numerical}
Butcher, J.~C. (2008).
\newblock {\em Numerical methods for ordinary differential equations}.
\newblock John Wiley \& Sons.

\bibitem[Del~Moral et~al., 2006]{del2006sequential}
Del~Moral, P., Doucet, A., and Jasra, A. (2006).
\newblock Sequential {M}onte {C}arlo samplers.
\newblock {\em Journal of the Royal Statistical Society: Series B (Statistical
  Methodology)}, 68(3):411--436.

\bibitem[Donnet and Samson, 2011]{donnet2011algorithm}
Donnet, S. and Samson, A. (2011).
\newblock {EM} algorithm coupled with particle filter for maximum likelihood
  parameter estimation of stochastic differential mixed-effects models.
\newblock Technical Report hal-00519576 v2, Universite Paris Descartes MAP5.

\bibitem[Fearnhead and Prangle, 2012]{fearnhead2012constructing}
Fearnhead, P. and Prangle, D. (2012).
\newblock Constructing summary statistics for approximate {B}ayesian
  computation: semi-automatic approximate {B}ayesian computation.
\newblock {\em Journal of the Royal Statistical Society: Series B (Statistical
  Methodology)}, 74(3):419--474.

\bibitem[Gelman et~al., 2014]{gelman2014understanding}
Gelman, A., Hwang, J., and Vehtari, A. (2014).
\newblock Understanding predictive information criteria for {B}ayesian models.
\newblock {\em Statistics and Computing}, 24(6):997--1016.

\bibitem[Golightly and Wilkinson, 2005]{golightly2005bayesian}
Golightly, A. and Wilkinson, D.~J. (2005).
\newblock Bayesian inference for stochastic kinetic models using a diffusion
  approximation.
\newblock {\em Biometrics}, 61(3):781--788.

\bibitem[Golightly and Wilkinson, 2006]{golightly2006bayesian}
Golightly, A. and Wilkinson, D.~J. (2006).
\newblock Bayesian sequential inference for nonlinear multivariate diffusions.
\newblock {\em Statistics and Computing}, 16(4):323--338.

\bibitem[Golightly and Wilkinson, 2008]{golightly2008bayesian}
Golightly, A. and Wilkinson, D.~J. (2008).
\newblock Bayesian inference for nonlinear multivariate diffusion models
  observed with error.
\newblock {\em Computational Statistics \& Data Analysis}, 52(3):1674--1693.

\bibitem[Hethcote, 2000]{hethcote2000mathematics}
Hethcote, H.~W. (2000).
\newblock The mathematics of infectious diseases.
\newblock {\em SIAM review}, 42(4):599--653.

\bibitem[Hoeting et~al., 2006]{hoeting2006model}
Hoeting, J.~A., Davis, R.~A., Merton, A.~A., and Thompson, S.~E. (2006).
\newblock Model selection for geostatistical models.
\newblock {\em Ecological Applications}, 16(1):87--98.

\bibitem[Hoeting et~al., 1999]{hoeting1999bayesian}
Hoeting, J.~A., Madigan, D., Raftery, A.~E., and Volinsky, C.~T. (1999).
\newblock Bayesian model averaging: a tutorial.
\newblock {\em Statistical Science}, 14(4):382--401.

\bibitem[Hooten and Wikle, 2008]{hooten2008hierarchical}
Hooten, M.~B. and Wikle, C.~K. (2008).
\newblock A hierarchical {B}ayesian non-linear spatio-temporal model for the
  spread of invasive species with application to the eurasian collared-dove.
\newblock {\em Environmental and Ecological Statistics}, 15(1):59--70.

\bibitem[Kass and Raftery, 1995]{kass1995bayes}
Kass, R.~E. and Raftery, A.~E. (1995).
\newblock Bayes factors.
\newblock {\em Journal of the American Statistical Association},
  90(430):773--795.

\bibitem[Kloeden and Platen, 1992]{kloeden1992numerical}
Kloeden, P. and Platen, E. (1992).
\newblock {\em Numerical Solution of Stochastic Differential Equations}.
\newblock Springer-Verlag.

\bibitem[Marin et~al., 2014]{marin2014}
Marin, J.-M., Pillai, N.~S., Robert, C.~P., and Rousseau, J. (2014).
\newblock Relevant statistics for {B}ayesian model choice.
\newblock {\em Journal of the Royal Statistical Society: Series B (Statistical
  Methodology)}, 76(5):833--859.

\bibitem[Marin et~al., 2012]{marin2012approximate}
Marin, J.-M., Pudlo, P., Robert, C.~P., and Ryder, R.~J. (2012).
\newblock Approximate {B}ayesian computational methods.
\newblock {\em Statistics and Computing}, 22(6):1167--1180.

\bibitem[Marjoram et~al., 2003]{marjoram2003markov}
Marjoram, P., Molitor, J., Plagnol, V., and Tavar{\'e}, S. (2003).
\newblock Markov chain {M}onte {C}arlo without likelihoods.
\newblock {\em Proceedings of the National Academy of Sciences},
  100(26):15324--15328.

\bibitem[Miller et~al., 2006]{miller2006dynamics}
Miller, M., Hobbs, N., and Tavener, S. (2006).
\newblock Dynamics of prion disease transmission in mule deer.
\newblock {\em Ecological Applications}, 16(6):2208--2214.

\bibitem[Miller and Williams, 2003]{miller2003prion}
Miller, M.~W. and Williams, E.~S. (2003).
\newblock Prion disease: horizontal prion transmission in mule deer.
\newblock {\em Nature}, 425(6953):35--36.

\bibitem[Pritchard et~al., 1999]{pritchard1999population}
Pritchard, J.~K., Seielstad, M.~T., Perez-Lezaun, A., and Feldman, M.~W.
  (1999).
\newblock Population growth of human {Y} chromosomes: a study of {Y} chromosome
  microsatellites.
\newblock {\em Molecular Biology and Evolution}, 16(12):1791--1798.

\bibitem[Robert et~al., 2011]{robert2011}
Robert, C.~P., Cornuet, J.-M., Marin, J.-M., and Pillai, N.~S. (2011).
\newblock Lack of confidence in approximate {B}ayesian computation model
  choice.
\newblock {\em Proceedings of the National Academy of Sciences},
  108(37):15112--15117.

\bibitem[Soetaert et~al., 2010]{desolve}
Soetaert, K., Petzoldt, T., and Setzer, R.~W. (2010).
\newblock Solving differential equations in {R}: Package desolve.
\newblock {\em Journal of Statistical Software}, 33(9):1--25.

\bibitem[Spiegelhalter et~al., 2002]{spiegelhalter2002bayesian}
Spiegelhalter, D.~J., Best, N.~G., Carlin, B.~P., and Van Der~Linde, A. (2002).
\newblock Bayesian measures of model complexity and fit.
\newblock {\em Journal of the Royal Statistical Society: Series B (Statistical
  Methodology)}, 64(4):583--639.

\bibitem[Sun et~al., 2015]{sun2015penalized}
Sun, L., Lee, C., and Hoeting, J.~A. (2015).
\newblock A penalized simulated maximum likelihood approach in parameter
  estimation for stochastic differential equations.
\newblock {\em Computational Statistics \& Data Analysis}, 84:54--67.

\bibitem[Tavare et~al., 1997]{tavare1997inferring}
Tavare, S., Balding, D.~J., Griffiths, R., and Donnelly, P. (1997).
\newblock Inferring coalescence times from {DNA} sequence data.
\newblock {\em Genetics}, 145(2):505--518.

\bibitem[Tenan et~al., 2014]{tenan2014bayesian}
Tenan, S., O'€™Hara, R.~B., Hendriks, I., and Tavecchia, G. (2014).
\newblock Bayesian model selection: The steepest mountain to climb.
\newblock {\em Ecological Modelling}, 283:62--69.

\bibitem[Toni et~al., 2009]{toni2009approximate}
Toni, T., Welch, D., Strelkowa, N., Ipsen, A., and Stumpf, M.~P. (2009).
\newblock Approximate {B}ayesian computation scheme for parameter inference and
  model selection in dynamical systems.
\newblock {\em Journal of the Royal Society Interface}, 6(31):187--202.

\bibitem[Watanabe, 2010]{watanabe2010asymptotic}
Watanabe, S. (2010).
\newblock Asymptotic equivalence of {B}ayes cross validation and widely
  applicable information criterion in singular learning theory.
\newblock {\em The Journal of Machine Learning Research}, 11:3571--3594.

\bibitem[Wikle, 2003]{wikle2003hierarchical}
Wikle, C.~K. (2003).
\newblock Hierarchical models in environmental science.
\newblock {\em International Statistical Review}, 71(2):181--199.

\end{thebibliography}
\label{lastpage}

\end{document}